\documentclass[%
 reprint,
 amsmath,amssymb,
 aps,
]{revtex4-2}
\usepackage[bookmarksnumbered=true,bookmarksopen=true]{hyperref}
 \hypersetup{colorlinks,%
             linkcolor=[rgb]{0,0.3,0.6}, %
             citecolor=[rgb]{0,0.3,0.6}, %
             urlcolor=[rgb]{0,0.3,0.6}}

\usepackage{amsmath}
\usepackage{graphicx}
\usepackage{dcolumn}
\usepackage{bm}

\def\ma{\mathcal}

\def\mb{\mathbf}
\def\fr{\frac}

\def\pd{\partial}

\def\tn{\tilde{\nu}}

\def\a{\mathrm{a}}
\def\b{\mathrm{b}}
\def\crm{\mathrm{c}}
\def\drm{\mathrm{d}}
\def\bgm{\begin{matrix}}
\def\edm{\end{matrix}}
\def\Ga{\Gamma}

\def\bgp{\begin{pmatrix}}
\def\edp{\end{pmatrix}}

\def\be{\begin{equation}}
\def\ee{\end{equation}}

\def\tilde{\widetilde}

\def\d{{\mathrm d}}

\def\bar{\overline}

\def\[{\bigl [}

\def\]{\bigr ]}

\def\la{\langle}

\def\ra{\rangle}

\def\tilde{\widetilde}

\def\bar{\overline}

\def\d{\mathrm{d}}
\def\a{\mathsf{a}}
\def\b{\mathsf{b}}

\def\Ga{\Gamma}

\def\mb{\mathbf}
\def\fr{\frac}

\def\be{\begin{equation}}
	\def\ee{\end{equation}}

\def\bar{\overline}
\def\pd{\partial}

\def\tn{\tilde{\nu}}
\def\vphi{\varphi}
\def\la{\langle}
\def\ra{\rangle}

\def\bgm{\begin{matrix}}
	\def\edm{\end{matrix}}
\newcommand{\te}{\text}
\newcommand{\blue}{\color[rgb]{0,.3,1}}

\begin{document}

\preprint{APS/123-QED}

\title{The cosmological collider signal in the non-BD initial states}

\author{Yuan Yin}%
 \email{yiny18@mails.tsinghua.edu.cn}
\affiliation{%
 Department of Physics, Tsinghua University, Beijing 100084, China
}%

\begin{abstract}
We investigate the cosmological collider (CC) signal arising from the tree-level exchange of a scalar spectator particle with a non-Bunch Davies (BD) initial state. We decompose the inflaton correlators into seed integrals, which we compute analytically by solving the bootstrap equations. We show that the non-BD initial state eliminates the Hubble scale Boltzmann suppression $e^{-\pi m /H}$ that usually affects the CC signal. Consequently, in this scenario, the CC can probe an energy scale much higher than the inflationary Hubble scale $H$. 
\end{abstract}

\maketitle
\section{Introduction}
Inflation is a widely accepted paradigm of the early universe that resolves various cosmological puzzles, such as the horizon problem \cite{Guth1981,Starobinsky1980}. Moreover, inflation predicts the cosmic non-Gaussianity (NG) \cite{Baumann2018} that offers a window into the deep UV physics that is inaccessible to any terrestrial particle collider. This paradigm, known as cosmological collider (CC) \cite{Chen2010,ArkaniHamed2015,Lee2016,Cui2022,Chen:2015lza, Chen:2016nrs, Chen:2016uwp, Chen:2016hrz, Lee:2016vti, An:2017hlx, An:2017rwo, Iyer:2017qzw, Kumar:2017ecc, Chen:2018cgg, Chua:2018dqh, Wu:2018lmx, Saito:2018omt, Li:2019ves, Lu:2019tjj, Liu:2019fag, Hook:2019zxa, Hook:2019vcn, Kumar:2018jxz, Kumar:2019ebj, Alexander:2019vtb, Wang:2019gok, Wang:2020uic, Li:2020xwr, Fan:2020xgh, Aoki:2020zbj, Lu:2021gso, Sou:2021juh, Lu:2021wxu, Pinol:2021aun, Cui:2021iie, Reece:2022soh, Qin:2022lva, Chen:2022vzh, Cabass:2022rhr, Cabass:2022oap, Niu:2022quw, Niu:2022fki, Aoki:2023tjm,Tong:2023krn,Qin:2023bjk}, has attracted broad attention in recent years. In CC, the inflaton is coupled to other spectator fields whose masses are above the inflationary Hubble scale. The interaction between the spectator fields and the inflaton leaves a distinctive imprint on the NG, which encodes the information of the mass and spin of the spectator fields. Despite its potential, CC is limited by the observational capability. The Planck satellite has constrained NG to a level of $f_{\text{NL}} \lesssim \ma{O}(10)$ \cite{Akrami2020}, and in the coming decades, SPHEREx can improve the sensitivity up to $f_{\text{NL}} \sim \ma{O}(1)$ \cite{SPHEREx:2014bgr}. Furthermore, through the probe of the 21cm line, we can eventually reach the resolution of $f_{\text{NL}} \sim \ma{O}(0.01)$ that touches the gravitational floor \cite{Munoz:2015eqa}. Therefore, inflation models that predict large CC signals become more interesting experimentally, as they can be tested by CC.

The CC is subject to two constraints. In the standard inflation models, the initial state is assumed to be the Bunch-Davies (BD) vacuum, and the particles observed at late times are solely produced by the cosmic expansion with an energy scale set by the inflationary Hubble scale $H$. Hence, the magnitude of the CC signals, often denoted as $f_{\te{NL}}$, is suppressed by the Hubble scale Boltzmann factor $f_{\te{NL}} \propto  e^{-\pi m/H}$. Moreover, one-particle states (here we consider scalar particles with $s = 0$) in de Sitter (dS) are divided into two series: the principal series for particles with $m > 3 H/2$ and the complementary series for particles with $0< m < 3H/2$. Only the particles in the principal series generate oscillatory signals. Therefore, the CC is insensitive to light particles. These two factors limit the detection window of the CC to the energy scale around the Hubble scale $H$.

In this article, we explore a class of inflation models that predict large cosmological collider signals for heavy spectator fields with mass much larger than the Hubble scale $H$. Unlike previous models that rely on an effective chemical potential \cite{Chen2018,Wang2020,Wang2020a,Bodas2021} to overcome the Hubble scale Boltzmann factor $e^{-\pi m /H}$ that suppresses the CC signals, we consider the scenario where inflation begin with the non BD state. 
The initial state we consider is the $\alpha$ vacuum \cite{GOLDSTEIN2003325}, which is related to the BD state \cite{Bunch1997} by a Bogoliubov transformation with a parameter $\alpha$. These initial states preserve the dS symmetry, and the case $\alpha = 0$ corresponds to the BD state. We show that in such inflation models, the cosmological collider  signals are free from the Hubble scale Boltzmann suppression. This implies a higher energy scale for the CC in this scenario. Moreover, this inflation model has a theoretical advantage of evading the swampland conjecture \cite{Ashoorioon:2018sqb}, which imposes constraints on low-energy effective theories compatible with quantum gravity. However, for a general $\alpha$ vacuum, the deviation from the BD state implies that it is excited. For a genuine $\alpha$ vacuum, the condition that the extra energy from the excited states is smaller than the vacuum energy imposes a strict constraint on $\alpha$. In this paper, we will adopt a flexible point of view as in \cite{Kanno2022}, where $\alpha$ depends on the momentum of the mode considered. Physically, such an initial state can be realized in inflation models such as warm inflation \cite{Tong2018}, where a thermal bath is generated during inflation due to rapid interactions.

In this paper, we adopt the Schwinger-Keldysh formalism \cite{Chen2017} to compute the inflaton correlator in de Sitter space. We show that the choice of the non-BD initial state modifies the Schwinger-Keldysh propagators of the inflaton. With the modified SK propagators, we can in principle calculate the non-BD inflaton correlators that contain the CC signals.

Apart from the phenomenological implication, the analytical properties of the inflaton correlators have attracted considerable attention in recent years, along with the interest in exploring the quantum field theory in dS. However, the dS correlators are still poorly understood. The computation of the inflaton correlators analytically faces several challenges. First, the dS propagator of massive field typically involves special functions that are difficult to integrate analytically. Second, the SK formalism introduces complex time-ordering into the calculation.

In recent years, some progress has been made in computing the inflaton correlators analytically. Two main methods are available: the Melin-Barnes (MB) formalism \cite{Sleight2020} and the bootstrap method \cite{ArkaniHamed2020,Qin2023,Qin2023a,Pimentel2022}. The MB formalism transforms the conformal time integral into an integral over the MB variables, which can be solved using the residue theorem. However, this method usually yields an infinite series expansion as the final result. The bootstrap method, which we adopt in this article, converts the integration problem into a differential equation. The inflaton correlator can be obtained analytically in certain limits, and these results serve as the boundary condition for the bootstrap equation. By solving this equation, we can obtain the closed-form expression for the inflaton correlator. These two methods have been used to compute the inflaton correlators involving tree exchange of scalar particles and gauge bosons under the BD initial condition. In this article, we extend the result to the non-BD case.

The rest of this paper is structured as follows. In section \ref{sec:TheModel}, we provide a brief overview of the model and the initial state. We also derive the mode function of the real scalar spectator field in non-BD scenario, which differs from the BD case by an additional negative frequency term. In section \ref{sec:CCS}, we briefly review the SK formalism and compute the non-BD propagators of the spectator field. Next, we demonstrate that the seed integrals are the basic components of the tree-level inflaton correlators. By calculating the CC signal contributions from the seed integrals, we can extract the general features of the tree-level inflaton correlators that are insensitive to the specific interactions. We employ the bootstrap method to calculate the tree-level exchange of the scalar particle. Section \ref{sec:conlusion} includes our concluding remark. The technical details of the calculation are presented in the Appendix \ref{sec:seedintegrallon}.

\section{The Model} \label{sec:TheModel}

In our model, the inflaton interacts with the scalar spectator field which has a non-BD initial state through the following action,

\be
S = S_\te{gravity} - \int \d^4 x \: \{ \ma{L}_\phi + \ma{L}_\sigma + \ma{L}_\te{int} \},
\ee

where $\ma{L}_\phi$, $\ma{L}_\sigma$, $\ma{L}_\te{int}$ denote the Lagrangian for the inflaton, the specator field, and the interaction respectively. The Lagrangian of the inflaton and takes the simplest form, which reads
\be
\ma{L}_\phi = \sqrt{-g}\Bigg[ \fr{1}{2}(\pd_\mu \phi)^2 + V(\phi)\Bigg],
\ee
and the $S_\te{gravity}$ is the Hilbert-Einstein action of general relativity that reads
\be
S_\te{gravity} = \fr{M_{\te{Pl}}^2}{2} \int \d^4 x \: \sqrt{-g} R.
\ee

We assume the inflaton has the BD initial sate, therefore, at leading order, our model give the conventional scale invariant power spectrum
\be
P_{\zeta}(k) = \fr{H^4}{(2\pi)^2 \dot{\phi}^2},
\ee
whose value was measured through CMB by PLANCK \cite{Akrami2020} to be $P_\zeta \sim 2\times 10^{-9}$. 

Most of the previous studies on inflation that started from a non-BD initial state focused on the scenario where the inflaton field had a $\alpha$-vacuum type initial state \cite{Xue2009,Akama2020,Holman2008,Ashoorioon2019,Kundu2014,Ashoorioon:2010xg,Meerburg2010,Agullo2011,Brahma2014,Agarwal2013,Ganc2012,Ganc2011,Emami2014,Shandera2010}. However, our scenario is different from theirs, as we assume the initial state of the inflaton is BD while the spectator field has the non-BD initial state. Nevertheless, we expect that our results can also be applied to their scenarios, which can be seen from the following discussion.

\subsection{Massive scalar in dS}
We assume the simplest scenario in which the spectator field is a scalar with a non-zero mass term. Therefore, the action for the free spectator field has the following expression,
\begin{align}
S_\sigma = & \:- \fr{1}{2} \int \d^4 x \: \sqrt{-g} \big[ g^{\mu\nu} \pd_\mu \sigma \pd_\nu \sigma + m^2 \sigma^2   \big] \\
  = & \: \fr{1}{2} \int \d\tau \d^3 \mb{x} \:  \big[ a^2 \sigma'^2 - a^2 (\pd_i \sigma)^2 - a^4 m^2 \sigma^2 \big]. \label{eq:actionsigma}
\end{align}
The equation of the motion can be derived by taking the variation of the action \eqref{eq:actionsigma}. The result is  
\be
\sigma''(\tau,\mb{x}) - \fr{2}{\tau}\sigma'(\tau,\mb{x}) - \pd_i^2 \sigma(\tau,\mb{x}) + \fr{m^2}{H^2 \tau^2} \sigma(\tau,\mb{x}) = 0. \label{eq:eomspecposition}
\ee
Subsequently, the equation of motion for each Fourier mode can be obtained by substituting $\sigma = \int \fr{\d^3 \mb{k}}{(2 \pi)^3} \sigma_\mb{k} e^{i \mb{k} \cdot \mb{x}}$ into \eqref{eq:eomspecposition}, which reads 
\be
\sigma_\mb{k}''(\tau) - \fr{2}{\tau}\sigma_\mb{k}(\tau) + \Big( k^2 + \fr{m^2}{H^2 \tau^2} \Big) \sigma_\mb{k}(\tau) = 0,
\ee
In the case of BD vacuum, we only keeps the solution with positive frequency as the mode function. Once the mode function is normalized by the Wronskian condition, it evantually takes the form of
\be
u(k,\tau) = \fr{\pi}{2}e^{-\tn \pi /2} H (-\tau)^{3/2} \te{H}_{i \tn}^{(1)}(- k \tau),
\ee
where $\tn  \equiv \sqrt{m^2/H^2-9/4}$. Canonically, the Fourier mode of the scalar field then can be quantized as 
\be
\sigma_{\mb{k}}(\tau) = u(k,\tau) b_{\mb{k}} + u^*(k,\tau) b_{-\mb{k}}^\dagger,
\ee
where $b_{\mb{k}}$ is the annihilation operator that annihilates the BD vacuum, $ b_{\mb{k}} |0\ra_{\te{BD}} = 0$. In this paper, we instead consider the non-BD initial state for the spectator field. Note that there are infinitely many states that satisfy the dS symmetry, therefore they can also become the candidates of the vacuum state. These vacuum, often called $\alpha$ vacuum, are linked to the BD vacuum by a Bogoliubov transformation as follows
\begin{eqnarray}
v(k,\tau) & = &  \text{cosh} \: \alpha \: u(k,\tau) + e^{i \phi} \te{sinh} \: \alpha \: u^*(k,\tau),\label{eq:bogoliubov0}\\
a_{\mb{k}} & = & \te{cosh} \: \alpha \: b_{\mb{k}} - e^{- i \phi} \te{sinh} \: \alpha \: b_{-\mb{k}}^\dagger, \label{eq:bogoliubov}
\end{eqnarray}
where $a_{\mb{k}}$ is the annihilation operator that annihilates the $\alpha$ vacuum, $ a_{\mb{k}} |0\ra_\alpha = 0$.

\subsection{Backreaction constraints}
The $\alpha$-vacuum initial state poses a  challenge for inflation models, as it deviates from the Bunch-Davies (BD) state and implies an excited state. To prevent the energy of the $\alpha$-vacuum from spoiling inflation, a stringent constraint on $\alpha$ is required.
Following  \cite{Kanno2022}, we assume that the parameter $\alpha$ is accompanied with a cutoff, that is, the $\alpha$ vacuum applies only to modes with $- k\tau < z_\Lambda$. This allows us to avoid the problem of backreaction in a large parameter space. The energy density can be evaluated as 
\be
\ma{E} \sim \int_{|\mb{k}| < z_\Lambda} \d^3 \mb{k} \: \sqrt{\mb{k}^2 + m^2} \: \te{sinh} \: 2\alpha . \label{eq:energyofspectator}
\ee
As argued, the inflation requires that the vacuum energy dominates. Therefore we reach the first constraint,
\be
\ma{E} \lesssim 3 M_{\te{Pl}}^2 H^2. \label{eq:constraint1}
\ee
Using \eqref{eq:energyofspectator}, the constraint \eqref{eq:constraint1} can be rewitten as 
\be
e^{2\alpha} \lesssim \fr{H^2 M_{\te{Pl}}^2}{z_\Lambda^4}. \label{eq:estimate1}
\ee
which is the constraint derived in \cite{Kanno2022}. The constraint on the parameter $\alpha$ depends on the Hubble scale. For example, we consider the scenario where $H/M_\te{Pl} \sim 10^{-10}$, that is, $H \sim 10^8 \: \te{GeV}$. As in \cite{Kanno2022}, we denotes $e^R \equiv z_\Lambda/H$, which is required to be much greater than unity to make our calculation valid, and it is severed as the second constraint. Then \eqref{eq:estimate1} implies that 
\be
e^\alpha \ll 10^{10}e^{-2R},
\ee
that is, $\alpha \ll 23 - 2R$. Therefore the constraint of the small backreaction still leaves us a large parameter space for $\alpha$ given that $R \gg 1$. In the next section, we will show that the cosmological collider signals obtain significant enhancement even with $\alpha \sim 1$.

\section{Cosmological Collider signals} \label{sec:CCS}
\subsection{Schwinger-Keldysh formalism} \label{sec:3A}
Conventionally, the cosmic correlator is calculated canonically by adopting the in-in formalism, the expectation value of an arbitrary operator $Q$ can be computed through,
\begin{align}
\la Q(\tau) \ra = & \la \Omega | \Big[ \bar{T} \te{exp} \left( 
i \int_{\tau_0}^\tau H_I (\tau) \d t  \right) Q(\tau)  \nonumber \\
& T \te{exp} \left( 
i \int_{\tau_0}^\tau H_I (\tau) \d t    \right)     \Big]     | \Omega \ra,
\end{align}
where $H_I$ is the Hamiltonian in the interaction picture. The correlator can be also calculated in the Lagrangian formalism as follows, 
\begin{align}
& \la \Omega | \varphi(\tau_f,\mb{x}_1) \cdots  \varphi(\tau_f,\mb{x}_n) | \Omega \ra = \int \ma{D} \varphi_+ \ma{D} \varphi_- \nonumber \\
& \times \varphi_+(\tau_f,\mb{x}_1)  \varphi_+(\tau_f,\mb{x}_n) \exp \left\{i \int_{\tau_0}^{\tau_f} \d^4 x \left( \ma{L}[\phi_+] - \ma{L}[\phi_-] \right)   \right\} \nonumber \\
&\qquad \times \prod_\mb{x} \delta \left[ \varphi_+(\tau_f,\mb{x}) - \varphi_-(\tau_f,\mb{x}) \right]. \label{eq:skcorrelator}
\end{align}

The calculation admits a diagrammatic approach, which analogus to the Feynman diagram. Similar to the flat spacetime, the basic ingredient of the calculation is the following SK propagators,
\begin{widetext}
\begin{align}
& D_{++}(k;\tau_1,\tau_2)(2\pi)^3\delta^{3}(\mb{k} + \mb{q}) 
=  \la 0_\alpha | T\{\sigma_\mb{k}(\tau_1) \sigma_\mb{q}(\tau_2)    \}  | 0_\alpha \ra, \\
& D_{--}(k;\tau_1,\tau_2)(2\pi)^3\delta^{3}(\mb{k} + \mb{q}) 
=  \la 0_\alpha | \bar{T}\{\sigma_\mb{k}(\tau_1) \sigma_\mb{q}(\tau_2)    \}  | 0_\alpha \ra, \\
& D_{+-}(k;\tau_1,\tau_2)(2\pi)^3\delta^{3}(\mb{k} + \mb{q})
=  \la 0_\alpha | \sigma_\mb{k}(\tau_1) \sigma_\mb{q}(\tau_2)  | 0_\alpha \ra, \\
& D_{-+}(k;\tau_1,\tau_2)(2\pi)^3\delta^{3}(\mb{k} + \mb{q}) 
=  \la 0_\alpha | \sigma_\mb{k}(\tau_1) \sigma_\mb{q}(\tau_2)  | 0_\alpha \ra. 
\end{align}
\end{widetext}
where $T\{ \cdots \}$ and $\bar{T}\{ \cdots \}$ denote the time ordering and anti time ordering. In contrast to the conventional BD initial state, the propagators here are obtained by taking the expectation value with respect to the $\alpha$-vacuum. Note that the four SK propagators are not independent, but related through 
\begin{widetext}
\begin{align}
    D_{++}(k;\tau_1,\tau_2) = & D_>(k;\tau_1,\tau_2)\theta(\tau_1-\tau_2) + D_<(k;\tau_1,\tau_2)\theta(\tau_2-\tau_1),\\
    D_{--}(k;\tau_1,\tau_2) = &D_<(k;\tau_1,\tau_2)\theta(\tau_1-\tau_2)+D_>(k;\tau_1,\tau_2)\theta(\tau_2-\tau_1),\\
    D_{+-}(k;\tau_1,\tau_2) = &D_<(k;\tau_1,\tau_2),\\
    D_{-+}(k;\tau_1,\tau_2) = &D_>(k;\tau_1,\tau_2).
\end{align}
\end{widetext}
where the Wightman function $D_>$ and $D_<$ are defined as

\begin{align}
D_>(k;\tau_1,\tau_2) = & \: v(k,\tau_1) v^*(k,\tau_2), \\
D_<(k;\tau_1,\tau_2) = & \: v^*(k,\tau_1) v(k,\tau_2),
\end{align}
where $v(k,\tau)$ is the mode function defined in \eqref{eq:bogoliubov0}. In terms of the BD mode function, the Wightman function can be rewritten as 
\begin{align}
& D_>(k;\tau_1,\tau_2) \nonumber \\
= & \: \te{cosh}^2 \: \alpha \: u(k,\tau_1)u^*(k,\tau_2) + \te{sinh}^2 \: \alpha \: u^*(k,\tau_1)u(k,\tau_2) \nonumber \\
 & \: + e^{-i \theta} \te{cosh} \: \alpha \: \te{sinh} \: \alpha  \:u(k,\tau_1)u(k,\tau_2) \nonumber \\ 
 & + e^{i \theta} \te{cosh} \: \alpha \: \te{sinh} \: \alpha  \: u^*(k,\tau_1)u^*(k,\tau_2), \label{eq:propagatoradvanced} \\
&  D_<(k;\tau_1,\tau_2) \nonumber \\
= & \: \te{cosh}^2 \: \alpha \: u^*(k,\tau_1)u(k,\tau_2) + \te{sinh}^2 \: \alpha \: u(k,\tau_1)u^*(k,\tau_2) \nonumber \\
 & \: + e^{i \theta} \te{cosh} \: \alpha \: \te{sinh} \: \alpha  \:u^*(k,\tau_1)u^*(k,\tau_2) + \nonumber \\ 
 & e^{-i \theta} \te{cosh} \: \alpha \: \te{sinh} \: \alpha  \: u(k,\tau_1)u(k,\tau_2). \label{eq:propagatorretard}
\end{align}
For clarity, we will decompose the propagator into four parts and analyze them respectively. We define the following four Wightman function,
\begin{align}
D_>^{(1)}(k;\tau_1,\tau_2) \equiv & \: u(k,\tau_1)u^*(k,\tau_2), \\
D_>^{(2)}(k;\tau_1,\tau_2) \equiv & \: u^*(k,\tau_1)u(k,\tau_2), \\
D_>^{(3)}(k;\tau_1,\tau_2) \equiv & \: e^{-i\theta} u(k,\tau_1)u(k,\tau_2), \\
D_>^{(4)}(k;\tau_1,\tau_2) \equiv & \: e^{i\theta} u^*(k,\tau_1)u^*(k,\tau_2). 
\end{align}
In the following, we will compute the contribution of the four parts of the propagator separately. For simplicity, we will take $\theta = 0$ from now on.

\subsection{Trispectrum}

\begin{figure}
 \centering
   \includegraphics[width=0.23\textwidth]{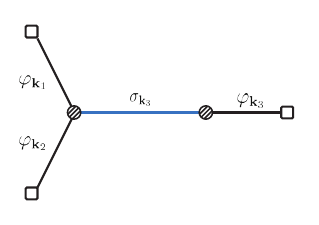}\label{fig:feynman3}
   \includegraphics[width=0.23\textwidth]{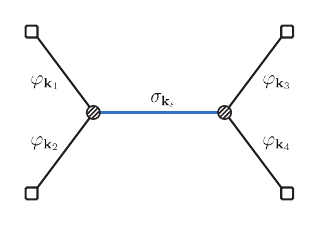}\label{fig:feynman4}
\caption{The Feynman diagrams of the 3-point correlator (\ref{eq:bispec}) and the 4-point correlator (\ref{eq:trispec}) following the Feynman rules listed in \cite{Chen2017}.}
\label{fig:feynman}
\end{figure}

We consider the simplest possibility of the interaction between the inflaton and the spectator and the inflaton which reserve the shift symmetry of the inflaton, which is given by 
\be
\ma{L}_{\te{int}} = \fr{1}{2} \lambda a^2 (\vphi')^2\sigma + \kappa a^3 \vphi' \sigma, \label{eq:typeofinteraction}
\ee
where $\vphi$ denotes the inflaton fluctuation. Such interaction generates the tree level 4-point correlator that can be computed using the SK formalism, and takes the following form
\begin{align}
& \la \vphi_{\mb{k}_1} \vphi_{\mb{k}_2} \vphi_{\mb{k}_3} \vphi_{\mb{k}_4} \ra'_{\sigma} \nonumber \\
= & -\lambda^2 \sum_{\a,\b = \pm} \a\b \int_{-k_s/z_\Lambda}^0 \fr{\d\tau_1}{(-H \tau_1)^2}\fr{\d\tau_2}{(-H \tau_2)^2} \nonumber \\
& \times G_\a'(k_1,\tau_1)G_\a'(k_2,\tau_1) G_\b'(k_3,\tau_2)G_\b'(k_4,\tau_2)\nonumber \\
& \times D_{\a\b}(k_s;\tau_1,\tau_2), \label{eq:trispec}
\end{align}
where $G_\a = \fr{H^2}{2k^3} (1 - i \a k \tau)e^{i \a k \tau}  $ is the inflaton propagator. As we previously argued, we have to impose a cutoff on the integration \eqref{eq:trispec}, implying the  $k$-dependence of $\alpha$. To avoid the unphysical and cumbersome hard cutoff on the $\alpha$-vacuum contribution of the correlator, we adopt a soft cutoff instead. We define the seed integral with a soft cutoff as follows:
\begin{align}
& \la \vphi_{\mb{k}_1} \vphi_{\mb{k}_2} \vphi_{\mb{k}_3} \vphi_{\mb{k}_4} \ra'_{\sigma} \nonumber \\
= & - \fr{H^4 \lambda^2}{16 k_1 k_2 k_3 k_4} \sum_{\a,\b = \pm}  \a\b \int^0_{-\infty} \d \tau_1 \d \tau_2 e^{i \a k_{12\a} \tau_1 + i \b k_{34\b} \tau_2} \nonumber \\
& \qquad \qquad \qquad \qquad \times D_{\a\b}(k_s;\tau_1,\tau_2) , \label{eq:integrationexample}
\end{align}
where we have denoted $k_{12\a} \equiv k_{12} - i \a H k_s/z_\Lambda$ and $k_{34\b} \equiv k_{34} - i \b H k_s/z_\Lambda$ with $k_{12} \equiv |\mb{k}_1| + |\mb{k}_2| $ and $k_{34} \equiv |\mb{k}_3| + |\mb{k}_4| $. Note that the imaginary part of the new external momentum suppresses the contribution from the region $-k_s\tau_{1,2} \gg z_\Lambda / H$. The integration appears in \eqref{eq:integrationexample} is generally encountered in the tree level calculation of the inflaton 4-point correlator and is not restricted to the interaction \eqref{eq:typeofinteraction}. Therefore it motivate us to analyze the following seed integrals instead, 
\begin{align}
\ma{I}^{(n),p_1p_2}_{\a\b} \equiv - \a\b & k_s^{5 + p_1 + p_2}  \int_{-\infty}^{0} \d \tau_1  \d \tau_2  \: (-\tau_1)^{p_1} (-\tau_2)^{p_2}  \nonumber \\
& \times e^{i \a k_{12\a} \tau_1 + i \b k_{34\b} \tau_2} D^{(n)}_{\a\b}(k_s;\tau_1,\tau_2). \label{eq:BDcoldseed}
\end{align}
The seed integrals determine the tree level inflaton correlator of any interaction, and we can infer its properties from them. The seed $\ma{I}^{(1),p_1p_2}_{\a\b}$ is the integral that will be encountered in BD calculation, while others are non-BD contribution. For instance, in term of the seed integrals, the 4-point correlator \eqref{eq:integrationexample} can be written as
\begin{align}
& \la \vphi_{\mb{k}_1} \vphi_{\mb{k}_2} \vphi_{\mb{k}_3} \vphi_{\mb{k}_4} \ra'_{\sigma} \nonumber \\
= & \: \fr{H^4 \lambda^2}{16 k_1 k_2 k_3 k_4 k_s^5}  \sum_{\a,\b = \pm} \Big\{ \te{cosh}^2\: \alpha \:  \ma{I}^{(1),00}_{\a\b} + \te{sinh}^2\: \alpha \:  \ma{I}^{(2),00}_{\a\b} \nonumber \\
& + \te{cosh} \: \alpha \: \te{sinh} \: \alpha \: \ma{I}^{(3),00}_{\a\b} + \te{cosh} \: \alpha \: \te{sinh} \: \alpha \: \ma{I}^{(4),00}_{\a\b} \Big\}. \label{eq:computethetri}
\end{align}
We obtain these seed integrals by solving the bootstrap equation, which we derive in detail in Appendix \ref{sec:seedintegrallon}. The general solution of the bootstrap equation has the form: 

\begin{align}
\ma{I}^{(n),p_1p_2}_{\a\b}(u_{1\a},u_{2\b}) = & \ma{V}^{(n),p_1p_2}_{\a\b}(u_{1\a},u_{2\b}) \nonumber \\
& +  \sum_{\crm,\drm = \pm} \alpha_{\a\b|\crm\drm}^{(n),p_1p_2}\:  \ma{Y}_\a^{p_1}(u_{1\a})\ma{Y}_\b^{p_2}(u_{2\b}), \nonumber \\
& \qquad \qquad \qquad(n=1,\dots,4) \label{eq:nonBDthermalseedgeneralsol}
\end{align}
where $u_{1\a} \equiv u_1(k_{12} \to k_{12\a})$, $u_{2\b} \equiv u_2(k_{34} \to k_{34\b})$ (We  also define $u_{1,2} \equiv 2r_{1,2}/(r_{1,2} + 1)$ and $r_{1,2} \equiv k_s/k_{12,34}$) and  $\ma{Y}_{\pm}^{p_1}(u)$ are the homogeneous solution of the bootstrapping equations that are defined as follows:
\begin{align}
\ma{Y}_{\pm}^{p}(u) = & 2^{\mp i \tn}\left(\fr{u}{2}\right)^{5/2 + p \pm i \tn} \Gamma\left( 5/2 + p \pm i \tn\right)\Gamma\left( \mp i \tn \right) \nonumber \\
& \times {}_2\text{F}_1 \left[\bgm 5/2 + p \pm i \tn , 1/2 \pm i \tn  \\ 1 \pm 2 i \tn\edm\middle|\, u \right],
\end{align}
where ${}_2\text{F}_1\left[\bgm 5/2 + p_1 \pm i \tn , 1/2 \pm i \tn  \\ 1 \pm 2 i \tn\edm\middle|\, u \right]$ denotes the hypergeometric function. $\ma{V}^{(n),p_1p_2}_{\a\b}$ denotes the particular solution and the $\alpha$ coefficients are the integration constants of the bootstrap equation. Following Appendix \ref{sec:seedintegrallon}, we can further decompose (\ref{eq:nonBDthermalseedgeneralsol}) into,
\begin{align}
\ma{I}^{(n),p_1p_2}_{\a\b|\text{bg}}(u_{1\a},u_{2\b}) \equiv  \: \ma{V}^{(n),p_1p_2}_{\a\b}(u_{1\a},u_{2\b}), \label{eq:defofbg}
\end{align}
that exclusively contribute to the background, and 
\begin{align}
& \ma{I}^{(n),p_1p_2}_{\a\b|\text{signal}}(u_{1\a},u_{2\b}) \nonumber \\
\equiv & \sum_{\crm,\drm = \pm} \alpha_{\a\b|\crm\drm}^{(n),p_1p_2}\:  \ma{Y}_\a^{p_1}(u_{1\a})\ma{Y}_\b^{p_2}(u_{2\b}), \label{eq:defofsig}
\end{align}
which contain all the contributions to the oscillatory signal. We can uniquely determine the particular solutions and the integration constants by imposing the collapsed limit result as the boundary condition (see Appendix \ref{sec:seedintegrallon}), results are,

For $\ma{I}^{(1)}_{\a\b}$:
\begin{align}
\alpha_{+-|\crm\drm}^{(1),p_1p_2} = & \alpha_{-+|\crm\drm}^{(1),p_1p_2*} = \fr{H^2 e^{-i \pi (p_1 - p_2)/2}}{4\pi}, \\
\alpha_{++|+\pm}^{(1),p_1p_2} = & \alpha_{--|-\pm}^{(1),p_1p_2*} = \fr{i H^2 e^{-i \pi (p_1 + p_2)/2}e^{\pi \tn}}{4\pi}, \\
\alpha_{++|-\pm}^{(1),p_1p_2} = & \alpha_{--|+\pm}^{(1),p_1p_2*} = \fr{i H^2 e^{-i \pi (p_1 + p_2)/2}e^{-\pi \tn}}{4\pi}. 
\end{align}

For $\ma{I}^{(2)}_{\a\b}$:
\begin{align}
\alpha_{+-|\pm\pm}^{(2),p_1p_2} = & \alpha_{-+|\pm\pm}^{(2),p_1p_2*} = \fr{H^2 e^{-i \pi (p_1 - p_2)/2}}{4\pi}, \\
\alpha_{+-|\pm\mp}^{(2),p_1p_2} = & \alpha_{-+|\mp\pm}^{(2),p_1p_2*} = \fr{H^2 e^{-i \pi (p_1 - p_2)/2}\blue e^{\pm2\pi \tn}}{4\pi}, \\
\alpha_{++|\pm+}^{(2),p_1p_2} = & \alpha_{--|\pm-}^{(2),p_1p_2*} = \fr{i H^2 e^{-i \pi (p_1 + p_2)/2}e^{\pi \tn}}{4\pi}, \\
\alpha_{++|\pm-}^{(2),p_1p_2} = & \alpha_{--|\pm+}^{(2),p_1p_2*} = \fr{i H^2 e^{-i \pi (p_1 + p_2)/2}e^{-\pi \tn}}{4\pi}. 
\end{align}

For $\ma{I}^{(3)}_{\a\b}$:
\begin{align}
\alpha_{+-|\pm\pm}^{(3),p_1p_2} = & \alpha_{-+|\mp\mp}^{(3),p_1p_2*} = - \fr{ H^2 e^{i \pi (p_1 - p_2)/2 }e^{ \mp\pi \tn}}{4\pi}, \\
\alpha_{+-|\pm\mp}^{(3),p_1p_2} = & \alpha_{-+|\mp\pm}^{(3),p_1p_2*} = - \fr{H^2 e^{-i \pi (p_1 - p_2)/2}e^{\pm \pi \tn}}{4\pi}, \\
\alpha_{++|\pm\pm}^{(3),p_1p_2} = & \alpha_{--|\mp\mp}^{(3),p_1p_2*} = - \fr{i H^2 e^{-i \pi (p_1 + p_2)/2}\blue e^{\pm 2\pi \tn}}{4\pi}, \\
\alpha_{++|\pm\mp}^{(3),p_1p_2} = & \alpha_{--|\pm\mp}^{(3),p_1p_2*} = - \fr{i H^2 e^{-i \pi (p_1 + p_2)/2}}{4\pi}. 
\end{align}

\begin{figure}[t]
 \centering
   \includegraphics[width=0.34\textwidth]{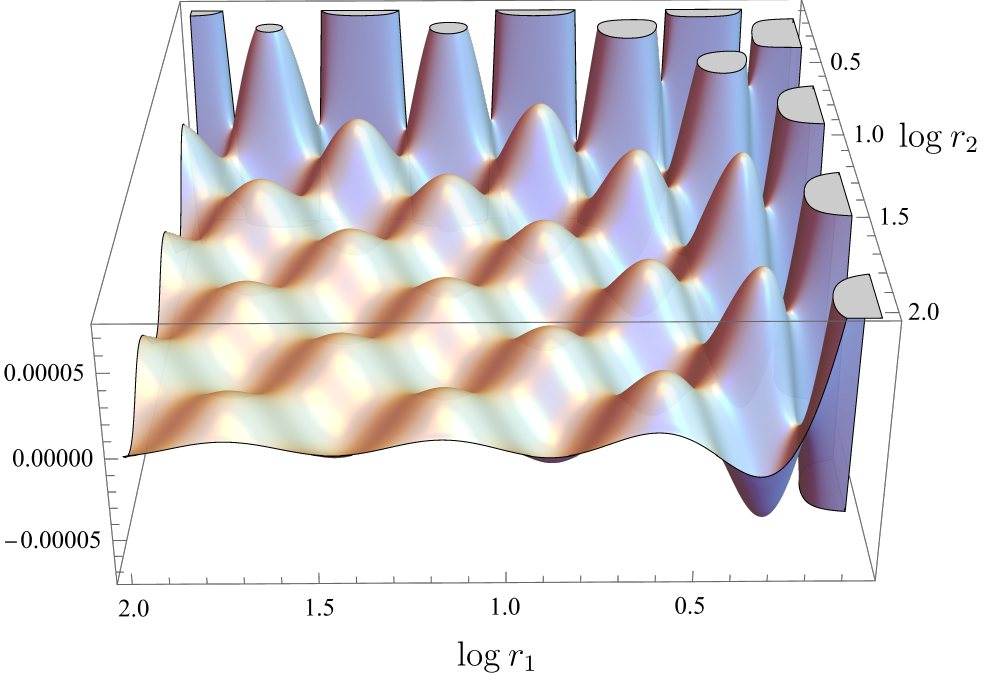}
   \includegraphics[width=0.34\textwidth]{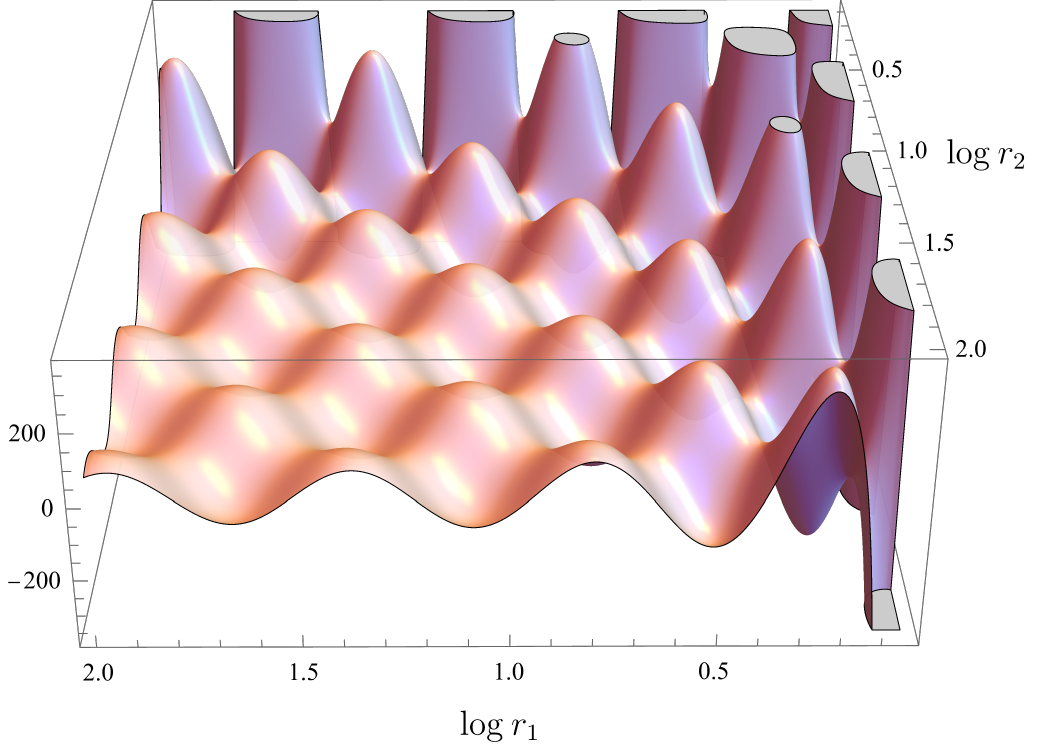}\\
   \includegraphics[width=0.34\textwidth]{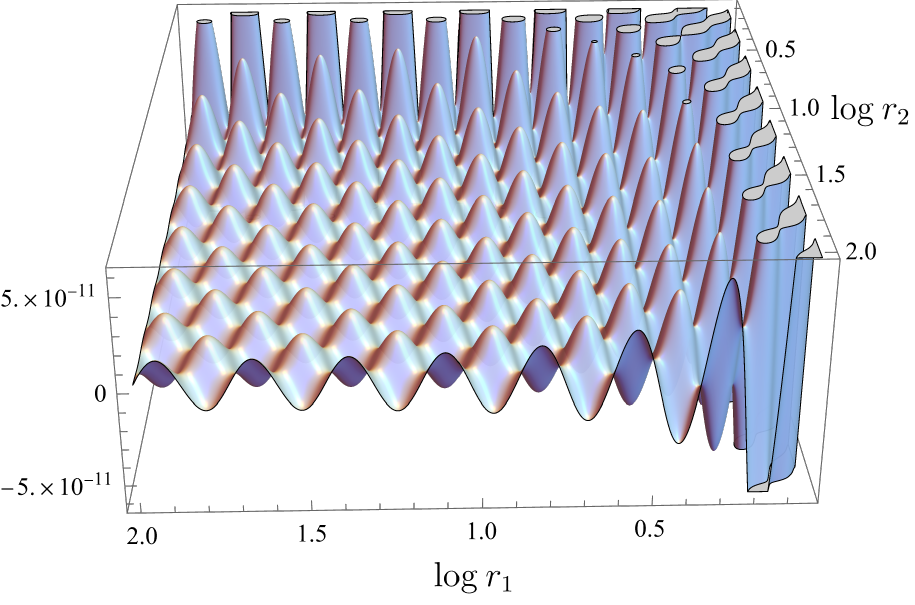}
   \includegraphics[width=0.34\textwidth]{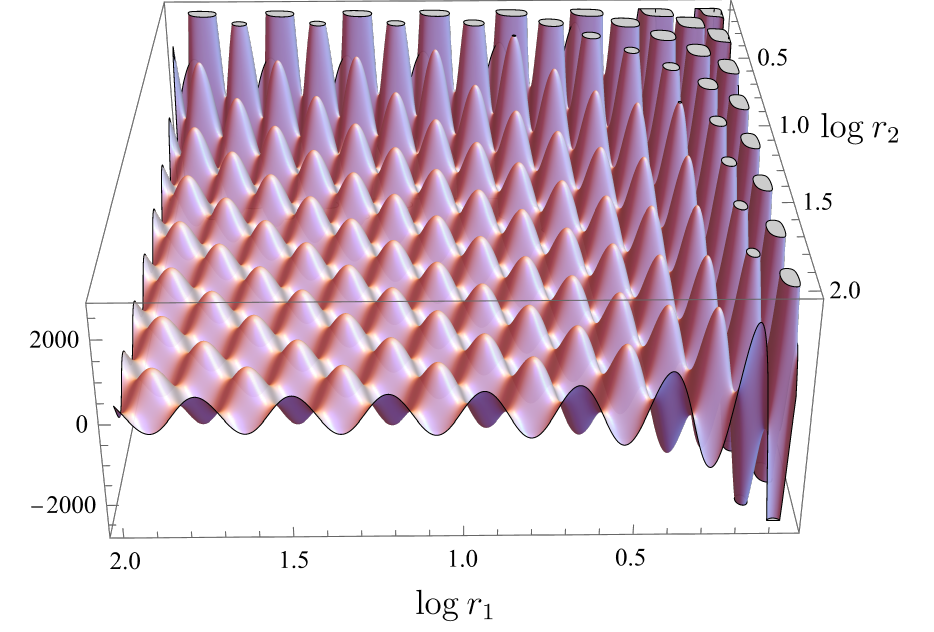}
\caption{The effect of different values of $\alpha$ and $m$ on the rescaled seed integral $\Big(\fr{2r_1}{r_1 + 1}\Big)^{-5/2}\Big(\fr{2r_2}{r_2 + 1}\Big)^{-5/2} \ma{I}^{00}_{\te{signal}}$. The blue and magenta curves correspond to $\alpha = 0$ and $\alpha = 1$ with $z_\Lambda/H = 20 $ , respectively. The first and second plots show the case of $m/H = 5$, while the third and fourth plots show the case of $m/H = 10$. The seed integral is a measure of the cosmological collider signal from the non-BD initial state.  We have taken $H = 1$ in these plots.}
\label{fig:4ptseed}
\end{figure}

For $\ma{I}^{(4)}_{\a\b}$:
\begin{align}
\alpha_{+-|\pm\pm}^{(4),p_1p_2} = & \alpha_{-+|\pm\pm}^{(4),p_1p_2*} = - \fr{H^2 e^{-i \pi (p_1 - p_2)/2} e^{\pm \pi \tn}}{4\pi}, \\
\alpha_{+-|\pm\mp}^{(4),p_1p_2} = & \alpha_{-+|\pm\mp}^{(4),p_1p_2*} = - \fr{H^2 e^{-i \pi (p_1 - p_2)/2} e^{\pm \pi \tn}}{4\pi}, \\
\alpha_{++|\pm\pm}^{(4),p_1p_2} = & \alpha_{--|\pm\pm}^{(4),p_1p_2*} = - \fr{i H^2 e^{-i \pi (p_1 + p_2)/2}}{4\pi}, \\
\alpha_{++|\pm\mp}^{(4),p_1p_2} = & \alpha_{--|\pm\mp}^{(4),p_1p_2*} = - \fr{i H^2 e^{-i \pi (p_1 + p_2)/2}}{4\pi}. 
\end{align}
We plot the signal contribution \eqref{eq:defofsig} from the seed integral in Figure \ref{fig:4ptseed} where we define $\ma{I}^{00}_{\te{signal}} \equiv \sum_{n,\a,\b} \ma{I}^{(n),00}_{\a\b|\te{signal}}$. We find that the seed integral is significantly enhanced by the non-BD contribution.  

\begin{figure}
 \centering
   \includegraphics[width=0.4\textwidth]{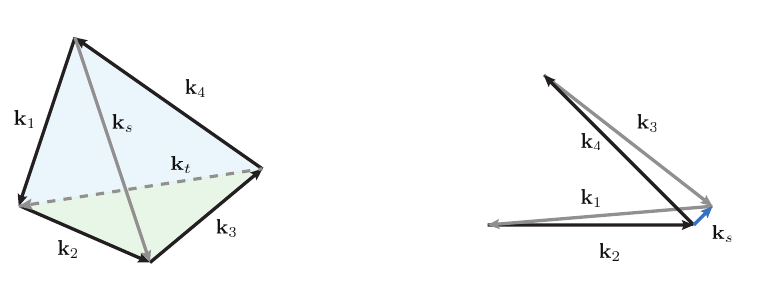}
\caption{Left panel: The momentum configuration for the 4-point correlator. Right panel: The collapsed limit configuration.}
\label{fig:4ptconfig}
\end{figure}

\paragraph{Signal Size Estimation}
We will estimate the size of the CC signal from the bispectrum in the follow. Conventionally, the trispectrum is captured through a dimenionless shape function defined as follows \cite{Chen2018a}, 
\begin{align}
& \ma{T}(k_1,k_2,k_3,k_4) \nonumber \\
= & \fr{1}{(2\pi)^6P_\zeta^3} \fr{H^4}{\dot{\phi}^4} \fr{(k_1k_2k_3k_4)^3}{K^3} \la \varphi_{\mb{k}_1}\varphi_{\mb{k}_2} \varphi_{\mb{k}_3} \varphi_{\mb{k}_4}  \ra'.
\end{align}
where $P_\zeta$ is the power spectrum.
The trispectrum is quantified by $t_{\text{NL}} \equiv |\ma{T}|$, which depends on the shape of the quadrangle formed by $\mb{k}_{1},\: \mb{k}_{2},\: \mb{k}_{3},\:\mb{k}_{4}$. Therefore, different values of $t_{\text{NL}}$ should be used for different quadrangle configurations. We focus on the case where one of the wave vectors is much smaller than the others, i.e., $k_s \ll k_{12},k_{34}$, i.e. the collapsed limit, see Figure \ref{fig:4ptconfig}. 

This limit exhibits non-analytic oscillations in the momentum ratios, which can be used to probe the presence of BSM particles from the CMB/LSS observations. We call this the CC signal and denote its amplitude by $t_{\text{NL}}^{(\text{osc})}$. From these definitions, it follows that $t_{\text{NL}}^{(\text{osc})} \sim \ma{T}(k_s \ll k_{12},k_{34})$ and can be estimated as \cite{Wang2020},
\be
t_{\text{NL}}^{(\text{osc})} \sim  \fr{1}{(2\pi)^2 P_\zeta} \la \varphi_{\mb{k}_1} \varphi_{\mb{k}_2} \varphi_{\mb{k}_3} \varphi_{\mb{k}_4} \ra'.
\ee
At the tree level, the 4-point correlators can be generally decomposed into seed integrals, which have signal and background parts as shown in (\ref{eq:defofbg}) and (\ref{eq:defofsig}). Therefore, the magnitude of $t_{\text{NL}}^{(\text{osc})}$ depends exclusively on the signal parts of the seed integrals (\ref{eq:defofsig}). We will focus on analyzing these seed integrals as proxies for the oscillatory signals. We will use the following asymptotic behavior of the gamma function:
\begin{align}
 \Gamma(a \pm i b) \simeq \sqrt{2 \pi} b^{a - 1/2} e^{-\pi b/2} e^{\pm i (b \; \text{log}b + \pi a /2)}, & \nonumber \\
 \qquad \qquad a,b \in \mathbb{R}, \quad b \gg 1. & \label{eq:gamma}
\end{align}

Note that even for moderately large values of $|b|$ greater than 1, this asymptotic behavior remains valid. 
For BD initial state,  $t_{\text{NL}}^{(\text{osc})}$ is suppressed by the factor $e^{-\pi m /H}$ at large mass which set the energy scale of the cosmological collider to $H$. This suppression can be found in the BD seed  $\ma{I}^{(1),p_1p_2}_{\a\b}$. In the collapsed limit $k_s \ll k_{12},k_{34}$, we can perform the following large mass expansion,
\begin{align}
\ma{Y}_{\pm}^{p_1}(u) \sim  \:\Ga\left( \fr{5}{2} + p_1  \pm i \tn \right) \Ga\left( \mp i \tn \right) \propto e^{-\pi m /H}, & \nonumber \\
\qquad  (m \gg H),&  \label{eq:modeasym}
\end{align}
where in the second step of (\ref{eq:modeasym}) we have applied (\ref{eq:gamma}). Among the coefficients in the BD seeds, the leading terms in the asymptotic expansion for large mass are given by
\begin{align}
\alpha_{++|+\pm}^{(1),p_1p_2} \sim \alpha_{--|-\pm}^{(1),p_1p_2}  \sim e^{\pi m/H},\label{eq:cofasym}
\end{align}
while the rest of the coefficients are further suppressed by the Boltzmann factor. Note that any physical calculation will require the summation over the Schwinger-Keldysh indices $\a,\b$. Hence, from the combination of equations (\ref{eq:modeasym}) and (\ref{eq:cofasym}), it is evident that the signal contribution from the BD seed is always governed by the factor $e^{-\pi m /H}$, which is consistent with our expectation. 

In contrast to the BD seeds, the leading coefficients from the non-BD seed are given by 
\begin{align}
\alpha_{+-|+-}^{(2),p_1p_2} = &  \alpha_{-+|-+}^{(2),p_1p_2}\sim e^{2\pi m /H},\label{eq:cofasymwa} \\
\alpha_{++|++}^{(3),p_1p_2} = & \alpha_{--|--}^{(3),p_1p_2}\sim e^{2\pi m /H}.\label{eq:cofsymwa2}
\end{align}
These coefficients in the non-BD seed cancel out the suppression factors in (\ref{eq:modeasym}). Hence, the non-BD seed $\ma{I}^{(2)}_{\a\b}$, $\ma{I}^{(3)}_{\a\b}$ does not suffer from the Hubble scale Boltzmann suppression $e^{-\pi m /H}$. Instead, its signal contribution is determined by the Bogoliubov coefficient $\te{sinh} \: \alpha $ in (\ref{eq:defofsig}).

To elucidate this phenomenon, we investigate the asymptotic behavior of the scalar mode function in the late-time regime and insert it into the 4-point seed integrals. \emph{The BD mode function} has the following form in the late-time expansion,
\begin{widetext}
\begin{align}
\lim_{k\to 0} u(k,\tau) = - i \sqrt{\fr{2}{\pi k^3}} H \left[ e^{\tn \pi/2}\Gamma( i \tn) \left( -\fr{k\tau}{2}\right)^{3/2 - i \tn} + e^{-\tn \pi/2}\Gamma( -i \tn) \left( -\fr{k\tau}{2}\right)^{3/2 + i \tn}  \right]. \label{eq:latetimemode}
\end{align}
\end{widetext}
On the other hand, as in Section \ref{sec:3A},
the non-time ordered BD propagators is defined as 
\begin{align}
D^{(1)}_{-+}(k;\tau_1,\tau_2) = & u(k,\tau_1)u^*(k,\tau_2), \label{eq:Dmp}\\
D^{(1)}_{+-}(k;\tau_1,\tau_2) = & u^*(k,\tau_1)u(k,\tau_2). \label{eq:Dpm}
\end{align}
Using \eqref{eq:latetimemode}, we observe that both (\ref{eq:Dmp}) and (\ref{eq:Dpm}) contain a term that is free from the exponential suppression in the late-time limit, which originates from the product of the first term in the bracket of \eqref{eq:latetimemode} and its complex conjugate, that is,
\be
D^{(1)}_{\a\b}(k;\tau_1,\tau_2) \supset \ma{O}(1)\:(\tau_1)^{3/2 - i \a \tn}(\tau_2)^{3/2 - i \b \tn},  \label{eq:proplt}
\ee
where $\a,\b$ take opposite sign. We take seed integral $\ma{I}^{(1)}_{\a\b}$ as an example, which is defined in (\ref{eq:BDcoldseed}) as 
\begin{align}
\ma{I}_{\a\b}^{(1),p_1p_2} \equiv & -\a\b k_s^{5+ p_1 + p_2} \int_{-\infty}^0 \d \tau_1 \d \tau_2 e^{i \a k_{12\a}\tau_1 + i \b k_{34\b}\tau_1} \nonumber \\
& \times (-\tau_1)^{p_1} (-\tau_2)^{p_2} D^{(1)}_{\a\b}(k_s, \tau_1,\tau_2).\label{eq:typicalint}
\end{align}
 Substituting (\ref{eq:proplt}) into (\ref{eq:typicalint}), we obtain a term in the seed integral that originates from (\ref{eq:proplt}) and is given by
\be
\ma{I}_{\a\b}^{(1),p_1p_2} \supset \ma{O}(1)  \Gamma (1 + i \tn) \Gamma(1 - i \tn).
\ee
From \eqref{eq:gamma}, this indicates that the seed integral is exponentially suppressed by the mass much larger than $H$, even though the propagator is not.  However, for non-BD initial condition, we will encounter the non-BD seed (here we consider $\ma{I}^{(2)}_{\a\b}$ as an example, the same holds for $\ma{I}^{(3)}_{\a\b}$) that has the following form,

\begin{align}
\ma{I}_{\a\b}^{(2),p_1p_2} \equiv -\a\b & k_s^{5+ p_1 + p_2} \int_{-\infty}^0 \d \tau_1 \d \tau_2 \:  e^{i \a k_{12\a}\tau_1 + i \b k_{34\b}\tau_2} \nonumber \\
& \times (-\tau_1)^{p_1} (-\tau_2)^{p_2} D^{(2)}_{\a\b}(k_s, \tau_1,\tau_2).\label{eq:typicalint2}
\end{align}
In contrast to (\ref{eq:proplt}), in the non-BD propagator, we have
\begin{align}
D^{(2)}_{\a\b}(k;\tau_1,\tau_2) = & D^{(1)}_{-\a-\b}(k;\tau_1,\tau_2) \nonumber \\
\supset &  \ma{O}(1)(\tau_1)^{3/2 + i \a \tn}(\tau_2)^{3/2 + i \b \tn}. \label{eq:warmpropexp}
\end{align}
It follows that, (\ref{eq:warmpropexp}) give rise to the term 
\be
\ma{I}_{\a\b}^{(2),p_1p_2} \supset \ma{O}(1) e^{\pi \tn} \Gamma (1 + i \tn) \Gamma(1 - i \tn).
\ee
which is free from the Hubble scale Boltzmann suppression. To conclude, the absence of the suppression in the non-BD seed is a result of the Schwinger-Kelydysh index mismatch, namely, the BD propagator that carries inverse SK indices $D^{(1)}_{-\a-\b}$ is matched with $e^{i \a k_{12}\tau_1 + i \b k_{34}\tau_1}$ in the non-BD seed (\ref{eq:typicalint2}). We remark that the late-time truncation of the propagator (\ref{eq:latetimemode}) is not applicable for computing the 3-point seed integrals, but this analysis still yields the correct exponential factor.

\subsection{Bispectrum}
The Feynman diagram in the left panel of Figure \ref{fig:feynman} leads to the CC signal in bispectrum and can be calculated using SK formalism. The expression for the 3-point correlator reads
\begin{align}
& \la \vphi_{\mb{k}_1} \vphi_{\mb{k}_2} \vphi_{\mb{k}_3}  \ra'_{\sigma} \nonumber \\
= & - \fr{H^4 \lambda \kappa}{8 k_1 k_2 k_3^4} \sum_{\a,\b = \pm}  \a\b \int^0_{-\infty} \d \tau_1 \d \tau_2 \: (-\tau_2)^{-2 } e^{i \a k_{12\a} \tau_1 + i \b k_{34\b} \tau_2} \nonumber \\
& \qquad \qquad \qquad \qquad \times D_{\a\b}(k_s;\tau_1,\tau_2) . \label{eq:bispec}
\end{align}
Analogue to the 4-point case, the tree level 3-point correlator can be generally constructed from the 3-point seed integral that takes the following form,
\begin{align}
\ma{I}^{(n),p_1p_2}_{\a\b} \equiv - \a\b & k_3^{5 + p_1 + p_2}  \int_{-\infty}^{0} \d \tau_1  \d \tau_2  \: (-\tau_1)^{p_1} (-\tau_2)^{p_2}  \nonumber \\
& \times e^{i \a k_{12\a} \tau_1 + i \b k_{3\b} \tau_2} D^{(n)}_{\a\b}(k_3;\tau_1,\tau_2) .
\end{align}
The 3-point seed can be obtained from the 4-point seed by applying the folded limit, $k_4 \to 0$. As in the trispectrum case, our result for the bispectrum is not restricted to the specific interaction \eqref{eq:typeofinteraction}, but rather it is applicable to any tree-level exchange process of the scalar particle, as it relies on seed integrals that can be combined in various ways.

As an example, we consider the seed integral appeared in (\ref{eq:bispec}), where $p_1 = 0,p_2 = -2$. We define a quantity that capture this seed integral’s contribution to the CC signal as
\be
\ma{I}^{0,-2}_{\text{signal}}(r) = \sum_{\a\b} \ma{I}^{0,-2}_{\a\b|\text{signal}}(u_\a(r),1_\b),
\ee
where $u_{\a}(r) \equiv u_{1\a}(k_s \to k_3)$, $1_\b \equiv u_{2\b}(k_4 \to 0)$ and $r \equiv k_3/k_{12}$.
\begin{figure}
 \centering
   \includegraphics[width=0.4\textwidth]{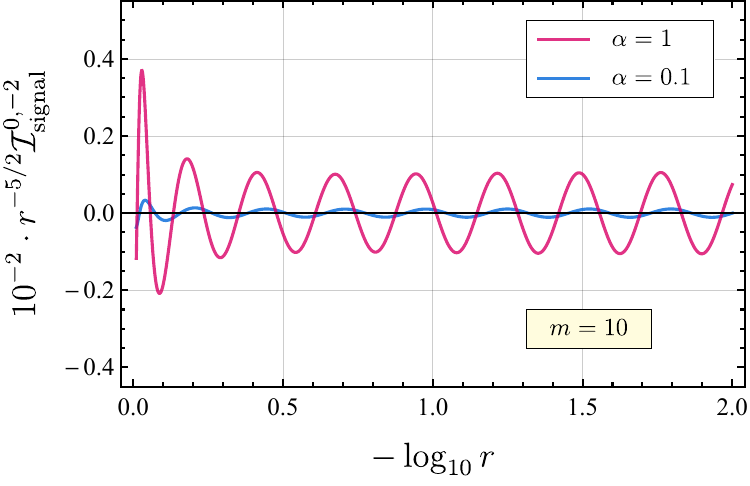}\\
   \includegraphics[width=0.4\textwidth]{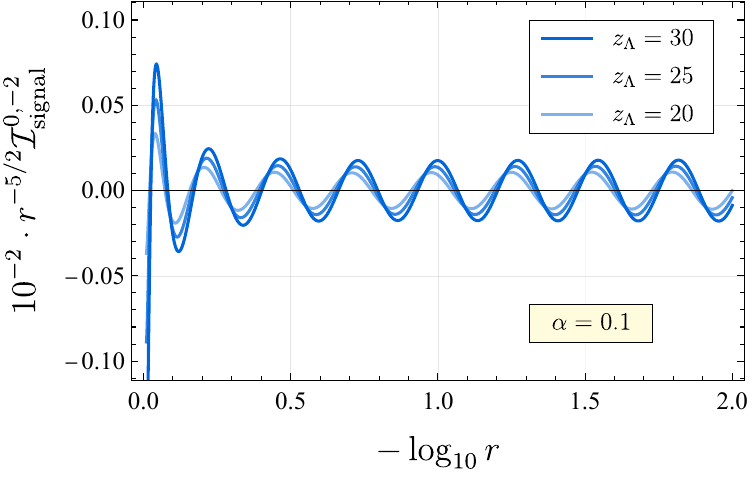}\\
   \includegraphics[width=0.4\textwidth]{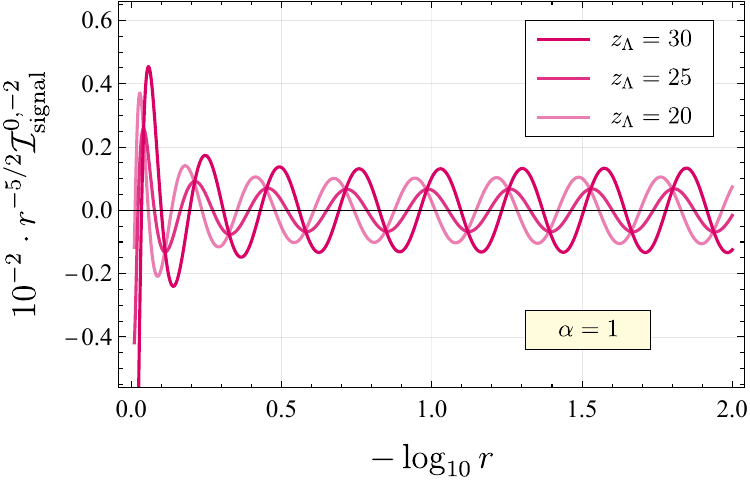}
\caption{ Upper panel: The seed integral as a function of the momentum configuration with $m/H =10$ and $z_\Lambda / H  = 20$. Middle panel and lower panel: the seed integral with different cutoff scale $z_\Lambda$. We have taken $H = 1$ in these plots.}
\label{fig:3ptsignal}
\end{figure}
We plot $\ma{I}^{0,-2}_{\text{signal}}$ in Figure \ref{fig:3ptsignal}. We see that the magnitude of the CC signal is substantially enhanced by the non-BD contribution.

\paragraph{Signal Size Estimation}
To quantify the oscillatory signal in the bispectrum, we can employ the same method that we utilized for the trispectrum. Analogous to to the trispectrum case, we can introduce a dimensionless shape function for the bispectrum as follows \cite{Wang2020},
\be
\ma{S}(k_1,k_2,k_3) \equiv -\fr{(k_1 k_2 k_3)^2}{(2\pi)^4 P_\zeta^2}\left(\fr{H}{\dot{\phi}}\right)^3 \la \varphi_{\mb{k}_1} \varphi_{\mb{k}_2} \varphi_{\mb{k}_3} \ra'.
\ee

\begin{figure}
 \centering
   \includegraphics[width=0.4\textwidth]{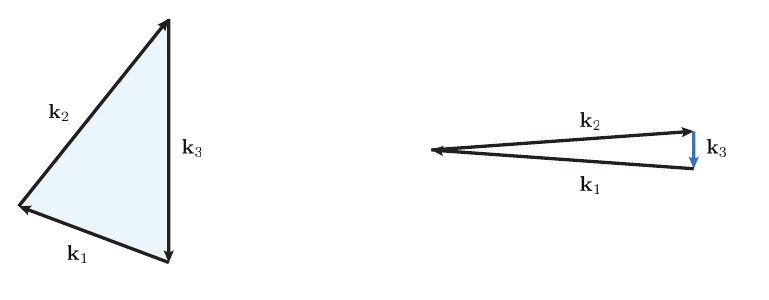}
\caption{Left panel: The momentum configuration for the 3-point correlator. Right panel: The squeezed limit configuration.}
\label{fig:3ptconfig}
\end{figure}

We can also introduce the analogue of $t_{\text{NL}}$ for the bispectrum as $f_{\text{NL}} \equiv |\ma{S}|$. For the bispectrum, the oscillatory signal appears in the squeezed limit, namely, $k_3 \ll k_{12}$ which is illustrated in the right panel of Figure \ref{fig:3ptconfig}.
Hence, we can estimate the magnitude of this signal by using a parameter $f_{\text{NL}}^{\text{osc}}$ that satisfies $f_{\text{NL}}^{\text{osc}} \sim |\ma{S}(k_3 \ll k_{12})|$ and can be estimated as, 
\be
f_{\text{NL}}^{(\text{osc})} \sim  \fr{1}{2\pi P_\zeta^{1/2}} \la \varphi_{\mb{k}_1} \varphi_{\mb{k}_2} \varphi_{\mb{k}_3} \ra'.
\ee
As argued, the tree level 3-point correlator can be decomposed into seed integrals.
Note that the 3-point seed integral can be obtained from the 4-point seed integral by applying the single folded limit, $k_4 \to 0$. Therefore, the non-BD contribution to the 3-point seed inherits the leading coefficients from the 4-point seed, given by \eqref{eq:cofasymwa} and \eqref{eq:cofsymwa2}. These coefficients cancel out the exponential suppression factor $e^{-\pi m/H}$ that appears in the mode function $\ma{Y}_{\pm}^{p}$.

\section{Conclusions and Discussions} \label{sec:conlusion}
Each collider has a characteristic energy scale, which also applies to the cosmological collider (CC). Usually, the energy scale of CC is limited by the fact that $f^{(\text{osc})}_{\text{NL}} \propto e^{-\pi m/H}$, which suppresses the signal from particles with mass much larger than the Hubble scale $H$. 

The energy scale of CC can be enhanced by considering models where the spectator field has an effective chemical potential \cite{Wang2020,Wang2020a,Chen2018}. These models still have a Hubble scale suppression, but it is offset by another exponential factor. We find that this suppression is absent if the spectator’s initial state is $\alpha$ vacuum, a class of vacua that preserve dS symmetry and are characterized by a parameter $\alpha$. We show that the magnitude of the oscillatory signal  depends on $\alpha$ in this scenario. We argue that CC can probe a wider range of particle masses in this case.

The inflaton correlator captures the cosmic non-gaussianity, which can be computed systematically by using the SK formalism. We demonstrate that the SK formalism can accommodate the non-BD initial state by replacing the SK propagators.

We have shown that the inflaton correlators can be expressed as a combination of seed integrals at tree level. By analyzing the seed integrals, we can reveal the general features of the CC signal that are independent of the interaction details. We  obtained the analytical expression for the full momentum configuration by solving the bootstrap equation with the boundary condition given by the collapsed limit. Our analysis reveals that the absence of the Hubble scale Boltzmann suppression originates from the non-BD part of the propagator. Since our result relies on the seed integrals, it is applicable to general tree level process and not restricted to a specific model. We emphasize that the $\alpha$ vacuum of the spectator field favors low-scale inflation. However, our result indicates that the cosmological collider can probe a much broader range of particle masses if the spectator has an $\alpha$ vacuum initial state.

Note that our calculation focus solely on tree level processes. In contrast, in some prominent inflation models, such as axion inflation \cite{Wang2020a}, features a spin-1 spectator field, and the dominant contribution comes from a bosonic loop. The generalization to loop calculation is intricate. Nevertheless, it has been shown that the loop process can also be tackled by the bootstrap method \cite{Xianyu:2022jwk}, we expect that our results can be extended to such process, which we will explore in future studies.

\begin{acknowledgments}
We thank Qi Chen for useful discussions.
\end{acknowledgments}

\appendix
\section{Scalar seed integrals}\label{sec:seedintegrallon}

In this appendix, we present a detailed derivation of the second seed integral $\ma{I}^{(2)}_{\a\b}$ appeared in \eqref{eq:computethetri}, following the method of \cite{Qin2023,Qin2023a}. The first seed integral $\ma{I}^{(1)}_{\a\b}$ is derived in \cite{Qin2023a}, and the rest of the seed integrals  $\ma{I}^{(3)}_{\a\b}$, $\ma{I}^{(4)}_{\a\b}$ can be derived in a similar way. The  seed integral $\ma{I}^{(2)}_{\a\b}$ is defined as
\begin{align}
& \ma{I}^{(2),p_1p_2}_{\a\b}(u_1,u_2;z_\Lambda) \nonumber \\
\equiv &    -\a\b k_s^{5+ p_1 + p_2} \int_{-\infty}^{0} \d \tau_1  \d \tau_2 (-\tau_1)^{p_1} (-\tau_2)^{p_2} \nonumber \\
& \qquad \qquad \times e^{i \a k_{12\a} \tau_1 + i \b k_{34\b} \tau_2} D^{(2)}_{\a\b}(k_s,\tau_1,\tau_2). \label{eq:defseed} 
\end{align}
The propagator $D^{(2)}_{\a\b}(k_s,\tau_1,\tau_2)$ satisfies the following equation of motion,
\begin{align}
(\tau_1^2 \pd_{\tau_1}^2 - 2\tau_1 \pd_{\tau_1} + k_s^2\tau_1^2 + m^2) D^{(2)}_{\pm\mp}(k_s;\tau_1,\tau_2) = & \: 0, \\
(\tau_1^2 \pd_{\tau_1}^2 - 2\tau_1 \pd_{\tau_1} + k_s^2\tau_1^2 + m^2) D^{(2)}_{\pm\pm}(k_s;\tau_1,\tau_2) & \nonumber \\
=  \pm i \tau_1^2 \tau_2^2 \delta(\tau_1 - \tau_2). &
\end{align}
To proceed, we define two new dimensionless variables as $z_{1\a} \equiv - k_{12\a}\tau_1$ and $z_{2\b} \equiv - k_{34\b} \tau_2$. In term of those variables, the equation of motion become,
\begin{align}
&(r_{1\pm}^2 \pd_{r_{1\pm}}^2 - 2r_{1\pm} \pd_{r_{1\pm}} + r_{1\pm}^2z_{1\pm}^2 + m^2) \nonumber \\
& \times \ma{D}^{(2)}_{\pm\mp}(r_{1\pm}z_{1\pm},r_{2\mp}z_{2\mp}) = 0, \label{eq:eomdpmmp}  \\
& (r_{1\pm}^2 \pd_{r_{1\pm}}^2 - 2r_{1\pm} \pd_{r_{1\pm}} + r_{1\pm}^2z_{1\pm}^2 + m^2) \ma{D}^{(2)}_{\pm\pm}(r_{1\pm} z_{1\pm},r_{2\pm}z_{2\pm}) \nonumber \\
& =  \pm i r_{1\pm}^2r_{2\pm}^2z_{1\pm}^2 z_{2\pm}^2 \delta(r_{1\pm}z_{1\pm} - r_{2\pm}z_{2\pm}), \label{eq:eomdpmpm} 
\end{align}
where $\ma{D}^{(2)}_{\a\b}(r_{1\a}z_{1\a} ,r_{2\b}z_{2\b}) \equiv k_s^3 D^{(2)}(k_s;\tau_1,\tau_2)$ and $r_{1\a} \equiv k_s/k_{12\a}$, $r_{2\b} \equiv k_s/k_{34\b}$. 
To derive the differential equation satisfied by the seed integrals, we apply the operator  $r_{1\a}^2 \pd_{r_{1\a}}^2 - 2r_{1\a} \pd_{r_{1\a}} + r_{1\a}^2z_{1\a}^2 + m^2$ to the propagator inside the seed integrals (\ref{eq:defseed}) and extract it from the integral. The main challenge arises from the term $r_{1\a}^2z_{1\a}^2$ , which requires repeated integration by parts to be pulled out of the integral. The first step of this process is given by
\begin{align}
& \int_0^{ k_{12\a}\cdot\infty} \d z_{1\a} z_{1\a}^{p} e^{- i \a z_{1\a}} z_{1\a} \ma{D}^{(2)}_{\a\b}(r_{1\a} z_{1\a}, r_{2\b} z_{2\b}) \nonumber \\
= \: & - i\a (p + 1 + r_{1\a} \pd_{r_{1\a}}) \nonumber \\
 & \times \int_0^{k_{12\a}\cdot\infty} \d z_{1\a} z_{1\a}^p e^{-i\a z_{1\a}} \ma{D}^{(2)}_{\a\b}(r_{1\a} z_{1\a}, r_{2\b} z_{2\b}). \label{eq:ibp1}
\end{align}
 Applying (\ref{eq:ibp1}) twice yields the following result,
\begin{align}
&\int_0^{ k_{12\a}\cdot\infty} \d z_{1\a} z_{1\a}^p e^{- i \a z_{1\a}} z_{1\a}^2 \ma{D}^{(2)}_{\a\b}(r_{1\a} z_{1\a},r_{2\b} z_{2\b}) \nonumber \\
= &  - (p + 2 + r_{1\a} \pd_{r_{1\a}}) (p + 1 + r_{1\a} \pd_{r_{1\a}}) \nonumber \\
& \times \int_0^{ k_{12\a}\cdot\infty} \d z z_{1\a}^p e^{- i \a z_{1\a}} \ma{D}^{(2)}_{\a\b}(r_{1\a} z_{1\a},r_{2\b} z_{2\b}), \label{eq:ibp2}
\end{align}
Therefore, by applying (\ref{eq:ibp2}), we can extract the term proportional to $r_{1\a}^2 z_{1\a}^2$ from the integral and obtain the following differential equation for the seed integral,
\begin{align}
& \Big[(r_{1\pm}^2 - r_{1\pm}^4)\pd_{r_{1\pm}}^2 - (2r_{1\pm} + (4 + 2 p_1)r_{1\pm}^3)\pd_{r_{1\pm}} \nonumber \\
& \qquad + ( (\tn^2 + 9/4 ) - (p_1 + 1)(p_1 + 2)r_{1\pm}^2 )    \Big] \nonumber \\
&\times [r_{1\pm}^{-1 - p_1}r_{2\mp}^{-1 - p_2}\ma{I}^{(2),p_1p_2}_{\pm\mp}(r_{1\pm},r_{2\mp})] = 0, \label{eq:bspmmpr} \\
& \Big[(r_{1\pm}^2 - r_{1\pm}^4)\pd_{r_{1\pm}}^2 - (2r_{1\pm} + (4 + 2 p_1)r_{1\pm}^3)\pd_{r_{1\pm}} \nonumber \\
& \qquad +  ( (\tn^2 + 9/4 ) - (p_1 + 1)(p_1 + 2)r_{1\pm}^2 )    \Big] \nonumber \\
&\times [r_{1\pm}^{-1 - p_1}r_{2\pm}^{-1 - p_2}\ma{I}^{(2),p_1p_2}_{\pm\pm}(r_{1\pm},r_{2\pm})] \nonumber \\
= &  e^{\pm i p_{12}\pi/2}\fr{r_{1\pm}^{4 + p_2}r_{2\pm}^{4 + p_1}}{(r_{1\pm} + r_{2\pm})^{5 + p_{12}}}\Gamma(5 + p_{12}), \label{eq:bspmpmr}
\end{align}
where $p_{12} \equiv p_1 + p_2$. In terms of $u_{1\a} = 2r_{1\a}/(r_{1\a} + 1)$. (\ref{eq:bspmmpr}) and (\ref{eq:bspmpmr}) can be rewritten as 
\begin{align}
&\Big\{ (u_{1\pm}^2 - u_{1\pm}^3)\pd_{u_{1\pm}}^2 - \Big[ (4 + 2p_1) - (1 + p_1)u_{1\pm}^2 \Big]\pd_{u_{1\pm}}\nonumber \\
&  + \Big[ \tn^2 + (p_1 + 5/2)^2 \Big] \Big\} \ma{I}^{(2),p_1p_2}_{\pm\mp}(u_{1\pm},u_{2\mp}) \nonumber \\
& = 0, \label{eq:bspmmpu}\\
& \Big\{ (u_{1\pm}^2 - u_{1\pm}^3)\pd_{u_{1\pm}}^2 - \Big[ (4 + 2p_1) - (1 + p_1)u_{1\pm}^2 \Big]\pd_{u_{1\pm}} \nonumber \\
& + \Big[ \tn^2 + (p_1 + 5/2)^2 \Big] \Big\} \ma{I}^{(2),p_1p_2}_{\pm\pm}(u_{1\pm},u_{2\pm}) \nonumber  \\
 = &  e^{\pm i p_{12} \pi/2} \Gamma (5 + p_{12})\nonumber \\
& \left( \fr{u_{1\pm}u_{2\pm}}{2(u_{1\pm} +u_{2\pm} - u_{1\pm} u_{2\pm})} \right)^{5 + p_{12}} . \label{eq:bspmpmu}
\end{align}
To cover the whole parameter space, we have to consider another ratio $r_{2\b} \equiv k_s/k_{34\b}$. This leads to two more differential equations for the propagator, which have the same form as (\ref{eq:bspmmpu}) and (\ref{eq:bspmpmu}). We omit them in the following discussion for brevity.
\paragraph{Boundary conditions}
To find the general solutions of (\ref{eq:bspmmpu}) and (\ref{eq:bspmpmu}) in the form of (\ref{eq:nonBDthermalseedgeneralsol}), we need to specify their boundary conditions. These can be obtained from the late-time expansion (\ref{eq:latetimemode}), which applies when  $k_s$ is much smaller than $k_{12}$ and $k_{34}$ (or equivalently, when $u_{1\a}$ and $u_{2\b}$ approach zero). Using this expansion, we can obtain the asymptotic expression for the BD scalar Wightmann function:
\begin{eqnarray}
    &\:&\lim_{k_s\to0}D^{(2)}_>(k_s;\tau_1,\tau_2)\nonumber \\
    & = &\frac{H^2}{4\tn}(\tau_1\tau_2)^{3/2}(\coth (\pi\tn)+1)\left(\frac{\tau_1}{\tau_2}\right)^{i\tn} \nonumber \\
    &\:&+\frac{H^2}{4\tn}(\tau_1\tau_2)^{3/2}\Gamma^2(-i\tn)\left(\frac{k_s^2\tau_1\tau_2}{4}\right)^{i\tn}\nonumber \\
    &\:&+(\tn\to-\tn).
\end{eqnarray}
By substituting the late-time expansion into the seed integral, we are able to reach an analytical expression at the squeezed limit. The results are,
\begin{widetext}
\begin{align}
\lim_{u_{1+},u_{2-}\to0}\ma{I}^{(2),p_1p_2}_{+-} = & \fr{H^2 e^{-i \bar{p}_{12}\pi/2} e^{3\pi\tn}}{4\tn} \cdot   \ma{G}^{p_1,p_2}_{+-}(\tn)       \left(\frac{k_{34-}}{k_{12+}}\right)^{i\tn} \left(  \fr{k_s^2}{k_{12+}k_{34-}} \right)^{5/2 + p_{12}}\nonumber \\
 & + \fr{H^2 e^{-i \bar{p}_{12}\pi/2} e^{-\pi\tn}}{4\tn} \cdot  \ma{G}^{p_1,p_2}_{-+}(\tn)    \left(\frac{k_{34-}}{k_{12+}}\right)^{-i\tn} \left(  \fr{k_s^2}{k_{12+}k_{34-}} \right)^{5/2 + p_{12}} \nonumber \\
  & +  \fr{H^2 4^{-i\tn} e^{-i \bar{p}_{12}\pi/2} e^{\pi\tn}}{4\tn} \fr{\Gamma(-i \tn)}{\Gamma(i \tn)}   \cdot   \ma{G}^{p_1,p_2}_{++}(\tn)    \left(\frac{k_s^2}{k_{12+}k_{34-}}\right)^{i\tn+p_{12}+\frac{5}{2}} \nonumber \\
  & + \fr{H^2 4^{i\tn}e^{-i \bar{p}_{12}\pi/2} e^{\pi\tn}}{4\tn}  \cdot \fr{\Gamma(i \tn)}{\Gamma(-i \tn)} \cdot \ma{G}^{p_1,p_2}_{--}(\tn)     \left(\frac{k_s^2}{k_{12+}k_{34-}}\right)^{-i\tn+p_{12}+\frac{5}{2}}, \label{eq:boundary3-pt+-}
\end{align}
where $\ma{G}^{p_1,p_2}_{\a\b}(\tn) \equiv (\coth (\pi  \tn )-1) \Gamma \left(p_1 +  i \a \tn +5/2\right) \Gamma \left(p_2 + i \b \tn + 5/2\right) $, and for the time ordered seed integral, we have
\begin{align}
\lim_{u_{1+}\gg u_{2-}\to0}\ma{I}^{(2),p_1p_2}_{++} = & \fr{iH^2 e^{-i p_{12}\pi/2} }{4\tn} \cdot \ma{A}_- (\tn)\ma{B}^{p_1,p_2}_{+-}(\tn)       \left(\frac{k_{34-}}{k_{12+}}\right)^{i\tn} \left(  \fr{k_s^2}{k_{12+}k_{34-}} \right)^{5/2 + p_{12}}\nonumber \\
 & + \fr{i H^2 e^{-i p_{12}\pi/2} }{4\tn} \cdot \ma{A}_+ (\tn) \ma{B}^{p_1,p_2}_{-+}(\tn)    \left(\frac{k_{34-}}{k_{12+}}\right)^{-i\tn} \left(  \fr{k_s^2}{k_{12+}k_{34-}} \right)^{5/2 + p_{12}} \nonumber \\
  & +  \fr{i H^2 4^{-i\tn} e^{-i p_{12}\pi/2}}{4\tn} \fr{\Gamma(-i \tn)}{\Gamma(i \tn)}   \cdot \ma{A}_+  (\tn)\ma{B}^{p_1,p_2}_{++}(\tn)    \left(\frac{k_s^2}{k_{12+}k_{34-}}\right)^{i\tn+p_{12}+\frac{5}{2}} \nonumber \\
  & + \fr{i H^2 4^{i\tn}e^{-i p_{12}\pi/2} e^{-2\pi\tn}}{4\tn}  \cdot \fr{\Gamma(i \tn)}{\Gamma(-i \tn)} \cdot \ma{A}_+ (\tn)\ma{B}^{p_1,p_2}_{--}(\tn)     \left(\frac{k_s^2}{k_{12+}k_{34-}}\right)^{-i\tn+p_{12}+\frac{5}{2}}, 
\end{align}
\end{widetext}
where $\ma{A}_{\pm}(\tn) \equiv \pm 1 + \text{coth}(\pi \tn)$, $\ma{B}^{p_1,p_2}_{\a\b}(\tn) \equiv \Gamma \left(p_1 +  i \a \tn +5/2\right) \Gamma \left(p_2 + i \b \tn + 5/2\right)$. The other seed integrals are obtained through
\begin{equation}\label{eq:boundary3-pt-+}
    \ma{I}^{(2),p_1p_2}_{-+} = \ma{I}^{(2),p_1p_2*}_{+-},
\end{equation}
\begin{equation}\label{eq:boundary3-pt--}
   \ma{I}^{(2),p_1p_2}_{--} = \ma{I}^{(2),p_1p_2*}_{++}.
\end{equation}
We note that the non-BD seed has a term that escapes the Hubble scale Boltzmann suppression, as shown by the Gamma function asymptotics in (\ref{eq:gamma}). (\ref{eq:boundary3-pt+-})-(\ref{eq:boundary3-pt--}) gives the boundary conditions for the bootstrap equations and fixes the integration constants in (\ref{eq:nonBDthermalseedgeneralsol}).

\paragraph{The particular solution}
The last piece of the solution is the particular solution. This can be find by power expanding both sides and matching the coefficients. The Taylor expansion of the inhomogenuous term in the bootstrap equation (\ref{eq:bspmpmu}) takes the form of, 
\begin{align}
& e^{\pm i p_{12} \pi/2} \Gamma (5 + p_{12})\left( \fr{u_{1\pm}u_{2\pm}}{2(u_{1\pm} + u_{2\pm} -u_{1\pm} u_{2\pm})} \right)^{5 + p_{12}} \nonumber \\
& = \: \sum_{n = 0}^{\infty} \ma{C}_{\pm} \left( \bgm -5 - p_{12} \\ n \edm \right)u_{1\pm}^{5+p_{12} +n} \left( \fr{1}{u_{2\pm}} - 1 \right)^n , \label{eq:taylorinhomo}
\end{align}
where $\left( \bgm -5 - p_{12} \\ n \edm \right)$ denotes the combinatorial number. The constant $\ma{C}_{\pm}$ is defined as,
\be
\ma{C}_{\pm} \equiv \fr{e^{\pm i p_{12} \pi/2} }{2^{5 + p_{12}}}\Gamma (5 + p_{12}).
\ee
(\ref{eq:taylorinhomo}) indicates us to take the following ansatz,
\begin{align}
& \ma{V}^{p_1,p_2}_{\pm}(u_{1\pm},u_{2\pm}) \nonumber \\
= & \: \sum_{m,n = 0}^{\infty} \ma{V}^{p_1,p_2}_{m,n|\pm} u_{1\pm}^{5 + p_{12} + m + n} \left( \fr{1}{u_{2\pm}} - 1 \right)^n .
\end{align}
Plug the ansatz into (\ref{eq:bspmpmu}) and matching the coefficients with (\ref{eq:taylorinhomo}) at each order, we reach the following recursion equations,
\begin{align}
& \left[(n + p_{12} + 5) (n + p_{12} - 2 p_1) + \tn^2 + \left( p_1 + \fr{5}{2} \right)^2 \right] \nonumber \\
& \times \ma{V}^{p_1,p_2}_{0,n|\pm} =  \:  \ma{C}_{\pm} \left( \bgm -5 - p_{12} \\ n \edm \right), \label{eq:final}
\end{align}
and
\be
\ma{V}^{p_1,p_2}_{m + 1,n|\pm} =  \ma{R}_{m,n}^{p_1,p_2} \ma{V}^{p_1,p_2}_{m,n|\pm},
\ee
where
\be
 \ma{R}_{m,n}^{p_1,p_2}  \equiv  \fr{\ma{M}^{p_1p_2}_{m,n}}{\ma{N}^{p_1p_2}_{m,n}},
\ee
with $\ma{M}^{p_1p_2}_{m,n} = (m + n + p_{12} + 5)(m + n + p_{2} + 3)\:$ and $\:\ma{N}^{p_1p_2}_{m,n} = (m + n + p_{12} + 6)(m + n + p_{12} - 2 p_1 + 1) + \tn^2 + \left( p_1 + \fr{5}{2} \right)^2 $. The solution of the recursion equations reads
\begin{align}
& \ma{V}^{p_1,p_2}_{m,n|\pm} \nonumber \\
= & \fr{ \left(n+p_2+3\right) \left(n+p_{12}+5\right)}{\tn ^2+\left(n+p_{12}+6\right) \left(n-2 p_1+p_{12}+1\right)+\left(p_1+\frac{5}{2}\right)^2}\nonumber \\
& \times \ma{C}_{\pm} \cdot
 \ma{F}_{m,n}^{p_1,p_2} \cdot \ma{V}^{p_1,p_2}_{0,n|\pm},
\end{align}
where 
\begin{align}
\ma{F}_{m,n}^{p_1,p_2} \equiv & \fr{\left(n+p_2+4\right)_{m-1} \left(n+p_{12}+6\right)_{m-1}}{\left(n-i \tn +p_2+\frac{9}{2}\right)_{m-1} \left(n+i \tn +p_2+\frac{9}{2}\right)_{m-1}} \nonumber \\
& \times \left( \bgm -5 - p_{12} \\ n \edm \right),
\end{align}
and $(\cdots)_m$ denotes the Pochhammer function with $\ma{V}^{p_1,p_2}_{0,n|\pm}$ given in \eqref{eq:final}. This completes the calculation for the particular solution. Note that there is no $(\cdots)^{i \tn}$ like non-analytical oscillation in the particular solution. Therefore our separation of signal and background in \eqref{eq:defofbg}, \eqref{eq:defofsig} is justified.

\nocite{*}

\bibliography{alpha}

\end{document}